\documentclass[12pt,showpacs,superscriptaddress,preprint]{revtex4}
\usepackage{epsf}
\usepackage{amsmath}

\def\beq{\begin{equation}}
\def\eeq{\end{equation}}
\def\bea{\begin{eqnarray}}
\def\eea{\end{eqnarray}}
\def\beqa{\begin{equation}\begin{array}{l}}
\def\eeqa{\end{array}\end{equation}}
\def\eqlab#1{\label{eq:#1}}

\def\Eqref#1{Eq.~(\ref{eq:#1})}

\def\Figref#1{Fig.~\ref{fig:#1}}

\def\sla#1{#1 \hspace{-2mm} \slash}
\def\slap{p \hspace{-1.6mm} \slash}
\def\slad{\partial \hspace{-2mm} \slash}

\def\half{\mbox{\small{$\frac{1}{2}$}}}

\def\third{\mbox{\small{$\frac{1}{3}$}}}

\def\barr{\left(\begin{array}{c}}
\def\earr{\end{array}\right)}
\def\bmat{\left(\begin{array}{cc}}
\def\emat{\end{array}\right)}
\def\al{\alpha}
\def\be{\beta}
\def\ga{\gamma} \def\Ga{{\it\Gamma}}
\def\de{\delta} \def\De{\Delta}
\def\veps{\varepsilon}  \def\eps{\epsilon}

\def\la{\lambda} \def\La{{\it\Lambda}}

 \def\Si{{\it\Sigma}}
\def\th{\theta}

\def\pa{\partial}

\def\pa{\partial}

\def\nn{\nonumber}

\def\mathscr{\mathcal}

\def\3d{3-D}

\begin{document}

\title{The lifetime
of unstable particles in electromagnetic fields}

\author{Daniele Binosi}
\affiliation{ECT* Trento,
Villa Tambosi, Villazzano,
I-38050 TN, Italy}

\author{Vladimir Pascalutsa }

\affiliation{ECT* Trento,
Villa Tambosi, Villazzano,
I-38050 TN, Italy}

\affiliation{Institut f\"ur Kernphysik, Johannes Gutenberg Universit\"at, Mainz D-55099, Germany}

\date{\today}

\begin{abstract}
We show that the electromagnetic moments of unstable particles (resonances) 
have an absorptive contribution which quantifies the change 
of the particle's lifetime in an external electromagnetic field.
To give an example we compute here 
the imaginary part of the magnetic moment
for the cases of the muon and the neutron
at leading order in the electroweak coupling.
We also consider an analogous effect for the strongly-decaying
$\De$(1232) resonance. The result
for the muon is 
Im$\,\mu = e\, G_F^2 m^3/768 \pi^3$, with $e$ the charge and 
$m$ the mass of the muon, $G_F$ the Fermi constant, which 
in an external magnetic field of $B$ Tesla give rise to
the relative change in the muon lifetime of $3\times 10^{-15}\, B$. 
For neutron the effect is of a similar magnitude. 
We speculate on the observable
implications of this effect. 
\end{abstract}


\maketitle

\section{Introduction}
The electromagnetic (e.m.) moments of a particle are among
the few fundamental quantities which describe the particle
properties and as such have thoroughly been studied. 
The most renowned examples
are the magnetic moments of
the electron and the muon which have been measured to unprecendented
accuracy and yielded a number of physical insights, 
see\cite{muon:gm2} for recent reviews.
What is far lesser known is that 
the e.m.\ moments of unstable particles are
complex numbers in general \cite{Avdeev:1998sx,Pascalutsa:2004je}. 
Their imaginary part reflects, of course,
the unstable nature of the particle, however,
the precise interpretation has been missing.  In this paper
we work out the relation, suggested first by
Holstein \cite{Barry06}, which should exist between 
the imaginary part of the magnetic moment and the effect
of an external magnetic field on particle's lifetime.

The argument for such a relation is very simple.
The (self-)energy  of the particle with a lifetime $\tau$
has an absorptive part, which has an
interpretation of the width $\Gamma=1/\tau$. The particle's 
magnetic moment $\vec{\mu}$ in the presence of magnetic field $\vec{B}$
induces the change in the energy: $-\vec{\mu}\cdot \vec{B}$. 
The latter contribution can then also change the width,
provided the magnetic moment has an absorptive part (Im$\,\mu\neq 0 $).  

The decay properties of unstable particles,
such as muon or neutron are extremely well 
studied and are widely used
for the precise determination of the
Standard Model  parameters\cite{PDG06,determ}. 
There are also a plethora of studies
of how these particles behave in e.m.\ fields.
A well-known example is the search for the neutron's 
electric dipole moment\cite{nEDM}. 
In view of these studies it is compelling to investigate 
how the decay properties of 
unstable particles may be affected by e.m.\ fields. 

The lifetime of unstable quantum-mechanical systems is known to
be affected by an e.m.\ field. Positronium provides a textbook 
example\cite{positronium}, where the effect arises due to the
admixture of para- ($S=0$) and ortho- ($S=1$) positronium states with
orbital momentum $l=0$ by the magnetic field interacting
with the magnetic moments of the constituents.
As the result, already in the field of $B=0.2$ Tesla, the lifetime
of ortho-positronium decreases by almost a factor of 2.

It is far from obvious 
how the same kind of an effect can arise for an elementary
unstable particle, e.g., the muon. 
The above-mentioned relation between the
imaginary part of the magnetic moment and the lifetime change
may, therefore,  
provide us with both an interpretation for the imaginary part of
the magnetic moment and the means to compute the effect
of the lifetime change.

In the following we examine in detail the case of the muon,
compute the leading contribution to Im$\, \mu$ and
the corresponding effect on the lifetime. Then we will briefly
discuss the cases of the neutron and of the $\De$-resonance.

\section{Muon decay ($\mu \to e\,\nu_e\nu_\mu $)} 
\begin{figure}[t,b]
\centerline{  \epsfxsize=5cm%
  \epsffile{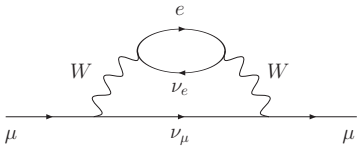} 
}
\caption{The muon self-energy contributing to its decay width.}
\label{fig:se}
\end{figure}

The leading contribution to the
muon decay  
width arises at two-loop  level, see \Figref{se}. 
For our purposes, the $W$ propagators in this graph can safely
be assumed to be static --- Fermi theory.
We also neglect the mass of the electron in the loops, since
it leads to an under-percent correction of $O(m_e/m)$;
here and in what follows, $m$ is the muon mass.  
The graphs with other Standard Model fermions (e.g., quarks)
in the loops need not to be considered here, 
because they cannot give any contribution to the muon width. 

Using dimensional regularization, we compute
this graph in $d=4-2\eps$ dimensions (in the limit $\eps\to 0^+$),\footnote{
Our conventions are: metric $(+,-,-,-)$, $\veps^{0123}=+1$,
$\ga_5=i\ga^0\ga^1\ga^2\ga^3$,
 $\ga$'s stand for Dirac matrices and their
totally-antisymmetric products:
$\ga^{\mu\nu}=\half [\ga^\mu,\,\ga^\nu]$,
$\ga^{\mu\nu\al}=\half \{\ga^{\mu\nu},\,\ga^\al\}$,
$\ga^{\mu\nu\al\be}=\half [\ga^{\mu\nu\al},\,\ga^\be]$.
}
\beq
\eqlab{selfen}
\Si(\slap) =  \frac{g^2}{8 M_W^4} 
i\!\int \frac{d^d k}{(2\pi)^d} \frac{2\ga_\mu (1-\ga_5)\, (\slap-\sla{k})\,
\ga_\nu }{(p-k)^2+i\veps}\,\Pi^{\mu\nu}(k).
\eeq
where $M_W$ is the $W$-boson mass, 
$g=|e|/\sin\th_W$ is the electroweak coupling 
related to the Fermi constant by $G_F/\sqrt{2} = g^2/8M_W^2$, 
$e$ is the charge, $\th_W$ is the Weinberg angle, and
\bea
\Pi^{\mu\nu}(k) & = & \frac{g^2}{8}\frac{d\, (d-2)}{(4\pi)^{d/2}(d-1)} \,
\frac{\Ga(\eps) \, \Ga(1-\eps)^2 }{\Ga(2-2\eps)} \nn\\
&\times & (-k^2)^{-\eps}\, \left(k^2 g^{\mu\nu} -k^\mu k^\nu
\right)
\eea
is the one-loop correction to the polarization tensor of the $W$ boson.  
The decay width can then be found as $\Gamma = - 2\, \mathrm{Im}
\Si(\slap = m)$.
A brief calculation shows that the self-energy has the following form:
\beq
\eqlab{genform}
\Si(\slap) = v(s) \, \slap \, (1-\ga_5) \, ,
\eeq
with $s=p^2$ and the scalar function $v$ given by:
\beq
v (s) =   -\frac{G_F^2\, s^2}{3(4\pi)^4} 
\left[\frac{1}{\eps}+\frac{21}{4} - 2\ga_E - 2\ln \frac{-s}{4\pi}
+O(\eps)\right],\\
\eeq
where $\ga_E=-\Ga'(1)$ is the Euler's constant. The absorptive part of this
function stems from the logarithm [$\ln (-s-i\veps) =\ln s - i\pi$,
for $s>0$]:
\beq
\eqlab{imv}
\mbox{Im} \, v (s) = -\frac{G_F^2\, s^2}{384\pi^3}\,.
\eeq
Terefore, the width is $\Gamma = -2m\,\mbox{Im} 
\,v(m^2)$, and the muon lifetime:  
\beq
\eqlab{oldresult}
\tau= 192\pi^3/(G_F^2 m^5)
\simeq 2.187\times 10^{-6}\, \mathrm{sec},
\eeq
This result is of course long-known due to the seminal work of
Feynman and Gell-Mann on Fermi theory\cite{Feynman:1958ty}. 
It is
in a percent agreement with 
the experimental value\cite{PDG06}:
\beq
\tau^{(\mathrm{exp})} = (2.19703\pm 0.00004) \, 10^{-6} \, \mathrm{sec},
\eeq
The discrepancy is due the neglect of the 
electron mass and some radiative corrections,
c.f.\cite{vanRitbergen:1998yd}.
We now
investigate the influence of the e.m.\ field
on the leading contribution given by \Eqref{oldresult}.

\begin{figure}[t,b]
\centerline{  \epsfxsize=5cm%
  \epsffile{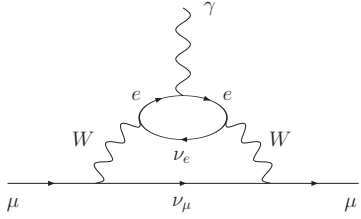} 
}
\caption{Electromagnetic correction to the muon decay.}
\label{fig:vert}
\end{figure}

Let us denote by $\Si (x,y; A_\mu)$
the self-energy in the presence of an external
e.m.\ field $A_\mu$. It is obtained
by minimal substitution ($\pa_\mu  \to \pa_\mu  -i e A_\mu$)
of the derivatives of all charged fields into the self-energy of
\Figref{se}.
Expanding in the e.m.\ coupling, we obtain:
\bea
\eqlab{seinA}
\Si [x,y; A_\mu] & = & \Si (i\sla{\pa}^{\,x})\, \de^4(x-y) \nn\\
 &+& \int \! dz\, \La^\mu (x,y;z)\,A_\mu(z)
+ O(e^2A^2),
\eea
where $\Si (i\sla{\pa})$ is the already computed self-energy
in the vacuum, 
while $\La$ is the e.m.\ vertex correction of \Figref{vert}, 
with static $W$'s.

Denoting $p$ $(p')$ the 4-momentum of the initial (final) muon
and assuming the on-shell situation 
($p^2={p'}^2=p\cdot p'=m^2$), the vertex correction
has in the momentum space the following general form:
\beq
\eqlab{vgenform}
\La^\mu(p',p)=e\left[F \, \ga^\mu + G\, \frac{(p+p')^\mu}{2m}
+F_A \, \ga^\mu\ga_5\right]\,,
\eeq
where $F$, $G$ and $F_A$ are complex numbers. 
Note that $e F/2m$ is the
correction to the magnetic moment, and $eF+eG$ is the correction
to the electric charge.
The Ward-Takahashi (WT) identity:
\beq
(p'-p)\cdot \La (p',p) = e\left[\Si(\slap) - \Si(\slap ')\right]
\eeq
 with the self-energy in \Eqref{genform} leads to the following conditions:
\beq
\eqlab{GIconds}
F+G=-v(m^2)-2m^2 v'(m^2),\,\,\, F_A=v(m^2)\,.
\eeq
Therefore, the term $F_A$
is in fact necessary by the e.m.\ gauge invariance. The $\ga_5$
terms, in both self-energy and the vertex, are shown to vanish when
summing over all the fermions in Standard Model\cite{Czarnecki:1996if}. 
However, this does not happen for the imaginary part because 
the heavier fermions do not contribute. 

The expression for the graph in \Figref{vert} is (in Fermi theory)
given by
\bea
&& \La^\mu(p',p) =  - \frac{e g^4}{64M_W^4} 
\int \frac{d^d k_1}{(2\pi)^d} \frac{d^d k_2}{(2\pi)^d}\, 2\ga_\al\, \frac{\sla{k}_2}{k_2^2}\,
\ga_\be (1-\ga_5)\nn\\
&&\,\,\times \,\frac{\mbox{Tr} \left[ 2\ga^\al (1-\ga_5)\, 
(\slap'-\sla{k}_1)\,\ga^\mu\,(\slap-\sla{k}_1)\,  
\ga^\be \, (\sla{k}_1 - \sla{k}_2 )\right]}{(k_1-k_2)^2\,(p-k_1)^2\, (p'-k_1)^2} .\nn\\
\eea
After a lengthy calculation we obtain the following result:
\beq
\eqlab{final}
\mbox{Im}\, F=\frac{G_F^2 m^4}{384\pi^3}, \,\,\, 
\mbox{Im}\,G=\frac{G_F^2 m^4}{96\pi^3},\,\,\, \mbox{Im}\,F_A=-\frac{G_F^2 m^4}{384\pi^3}\,,
\eeq
hence satisfying the gauge-invariance conditions~\Eqref{GIconds}, 
for Im$\,v$ given by \Eqref{imv}.

We would like to emphasize here
that, of course, not only the magnetic moment, but
also the charge operator receives an imaginary contribution, equal
to $e\,\mathrm{Im}(F+G)$. However, through
the WT identity, this contribution is completely fixed by the
momentum dependence of the self-energy, and therefore is not independent.
The same holds for $F_A$. We thus discuss only the effect
of the absorptive part of the magnetic moment, here given  by 
Im$\, \mu= e \,\mathrm{Im}\,F/2m = e G_F^2 m^3/768\pi^3$.

The energy of the magnetic moment interacting with the magnetic field
is equal to $-\mu B_z$, with $B_z$ being the projection of the field
along the muon spin. Then the total energy,
in the muon rest-frame, is given by: $ m-(i/2)\Gamma - \mu B_z$.
We thus deduce that the absorptive part yields
the following change in the muon width:
\beq
\De \Gamma = 2\, \mathrm{Im} \,\mu\, B_z = \frac{e}{2m}\,
 \frac{G_F^2 m^4}{192\pi^3}\, B_z\,,
\eeq 
while the change in the lifetime is $\De \tau = -  
(\De\Gamma/\Gamma)\,\tau $, for $\De\Gamma/\Gamma\ll 1$.

Given this result, we  conclude that
the positively-charged 
muons live shorter (longer) in a uniform magnetic field if their spin
is aligned along (against) the field. 
For the relative change in the width we find:
\beq
\frac{\left| \De \Gamma \right|}{\Gamma} 
= \frac{|e B_z|}{2m^2} 
\lesssim\, 3\times 10^{-15} \,B\,\,\mathrm{T}^{-1},
\eeq 
where $B$ is the strength of the field in Tesla.
Therefore, in moderate magnetic fields the change in 
the muon lifetime is
tiny, well beyond the present experimental accuracy (which is at
the ppm level). We will dwell on this more in the concluding
part of the paper, but for now we turn to a more technical issue.

It is interesting to observe that the result of \Eqref{final}, can simply
be obtained by the minimal substitution into \Eqref{genform},
rather than into the electron propagator in \Eqref{selfen}. 
To show this we go to coordinate
space and hence write the self-energy
 as $\Si(x,y) = \Si(i\slad)\, \de(x-y)$.
The minimal substitution to the first order in $e$
leads to the following vertex correction: 
\beq
\eqlab{minsub}
\tilde\La^\mu(x,y;z) = - \left.
\de/\de A_\mu (z)\, \Si(i\slad+e\sla{A}\,) \,\de(x-y)
\,\right|_{A=0}\,.
\eeq
Note that in general this is different from the vertex function in \Eqref{seinA},
since in the latter the minimal substitution is performed also in the internal lines.
The general form of 
\Eqref{vgenform}, of course, applies here as well, but now
the scalar functions are completely specified by the self-energy:
\beq
\eqlab{coin}
\tilde F=-v(m^2), \,\,\, \tilde G=-2m^2\,v'(m^2),\,\,\, \tilde F_A=v(m^2)\,.
\eeq
Substituting the explicit form of Im$\, v$, we see that
this method unambiguously leads to exactly the same result [\Eqref{final}]
as the full calculation. We emphasize though, that this method cannot
always work (see, e.g., Ref.\cite{Koch:2001ii}), 
as will also  be clear from the following examples.
Nevertheless, it is worthwhile to investigate this method further,
since knowing  whether it is applicable
{\it a priori}  can  enormously facilitate the calculations.

\section{Neutron decay and the $\De$-resonance}
We consider now the neutron $\beta$-decay.
Assuming exact $V-A$ interaction ($g_A=1$) and neglecting the electron mass
(but not the proton mass, $m_{\mathrm p}$), the corresponding two-loop self-energy 
can still be written in the form of \Eqref{genform}.
We introduce $\delta=(s-m^2_{\mathrm p})/2s$ 
and treat it as a small parameter, since
in the physical case  (where $s= m_{\mathrm n}^2$),  $\de \simeq 1.293\times 10^{-3}$.
A simple calculation then yields:
\beq
\mbox{Im}\ v(s) =-\frac{G_F^2\vert V_{\mathrm{ud}}\vert^2}{30\pi^3}\,s^2\, \delta^5,
\eeq
where $V_{\mathrm{ud}}$ is the relevant quark-mixing (CKM) matrix element.
We note in passing that this result leads to the lifetime
of $\tau_{\mathrm n}\approx 622$ sec, to be compared with the experimental
value of $886$ sec. This $30\,\%$ disagreement is largely
due to the fact that in reality 
the axial coupling $g_A$ deviates from 1. However,
for our order-of-magnitude estimate this discrepancy is unimportant. 

What is important
is that the derivative of the self-energy is enhanced by one power of
$\de$:
 \beq
\mbox{Im}\ v'(s)=-\frac{\left(G_F\vert V_{\mathrm{ud}}\vert\right)^2}{12\pi^3}\,s\,\delta^4\,.
\eeq
and this opens the possibility for the enhancement of the effect in the
lifetime. Namely,  the relative change in the neutron width then
goes as
\beq
\frac{\left| \De \Gamma_{\mathrm{n}} \right|}{\Gamma_{\mathrm{n}}} \sim
\frac{\mu_N |B_z|}{m_\mathrm{n}-m_\mathrm{p}}\lesssim\, 3\times 10^{-14} 
\,B\,\,\mathrm{T}^{-1},  
\eeq 
where $\mu_N\simeq 3.15\times 10^{-14}\ \mbox{MeV }\mbox{T}^{-1}$ is the nuclear magneton. A more precise analysis
of this effect for the neutron is beyond the scope of this paper.
We focus instead on the example of the $\De$-resonance, where
such an enhancement will be shown to be even more dramatic, at least
qualitatively. 

The $\De$ resonance
decays strongly into the pion and nucleon, $\De\to \pi N$,
and the corresponding self-energy, 
to leading order in chiral effective-field theory,
yields the following result for the absorptive part\cite{Pascalutsa:2004je}: 
\beq
\eqlab{width}
\mathrm{Im}\,\Si_\De(\slap) = -\mbox{$\frac{2}{3}$}\pi\la^3\,
C^2 \, (\al\, \slap + m_N) \,,
\eeq
where the isospin symmetry is assumed, 
e.g., $m_\mathrm{p}=m_\mathrm{n}=m_N$. The constant
$C=h_A m_\De/8\pi f_\pi \simeq 1.5$, where 
$h_A$ represents the $\pi N\De$ coupling and is fitted to the
empiracal width of the $\De$, $f_\pi\simeq 93$
MeV is the pion-decay constant,  and $m_\De = 1232$ MeV is the
$\De$ mass. 
For simplicity we neglect the pion
mass (i.e., take the chiral limit). Then, in \Eqref{width}, 
$\la = (s-m_N^2)/2s $, $\al = 1-\la $. For $s=m_\De$, $\la\approx 
(m_\De-m_N)/m_N \sim 1/3$ is a small parameter in the chiral 
effective-field theory with $\De$'s 
(see Ref.\cite{PVY07} for a recent review), 
and will so be treated here too.

\begin{figure}[t,b]
\centerline{  \epsfxsize=8cm%
  \epsffile{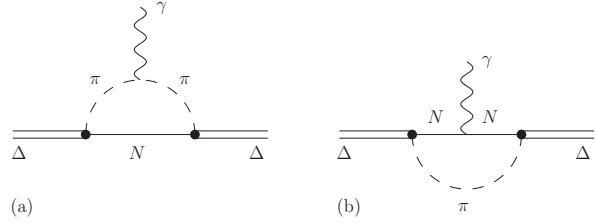} 
}
\caption{The leading chiral-loop correction to the magnetic 
moment of the $\De$. }
\label{fig:deltavert}
\end{figure}

The absorptive part of the magnetic dipole moment of the $\De$ arises
at this order from graphs in \Figref{deltavert}. These graphs, computed
in Ref.\cite{Pascalutsa:2004je}, in the chiral limit
yield the following result (upto $\la^4$ terms):
\bea
\mathrm{Im}\, F^{(a)} &=& 4\pi C^2 ( \la - 3 \la^2 +
\mbox{$\frac{43}{12}$} \la^3 )\,,\nn\\
\mathrm{Im}\, G^{(a)} &=& 4\pi C^2 ( -\la + 4 \la^2 -
\mbox{$\frac{71}{12}$} \la^3 )\,,\nn\\
 \mathrm{Im}\, F^{(b)} & = & 4\pi C^2 ( \la^2 +\third \la^3 )\,,\\
\mathrm{Im}\, G^{(b)} & =& -\mbox{$\frac{32}{3}$}\pi C^2  \la^3 \,,\nn    
\eea
where $F$ and $G$ correspond with the decomposition in \Eqref{vgenform},
with the superscript referring to the corresponding graphs in 
\Figref{deltavert};
$F_A$ is absent in this case, of course. 

First of all, we observe that this
result satisfies the WT conditions, \Eqref{GIconds}, for each of the 
four charge
states of the $\De$,
\bea
&& \De^{++}:  \,\mathrm{Im}\,[ F^{(a)}+G^{(a)}+F^{(b)}+G^{(b)}]
= -2\,\mathrm{Im}\,\Si_\De'\,,\nn\\ 
&& \De^{+}:  \, \mathrm{Im}\,[\third (F^{(a)}+G^{(a)})+
\mbox{$\frac{2}{3}$}(F^{(b)}+G^{(b)})]
= -\,\mathrm{Im}\,\Si_\De'\,,\nn\\ 
&& \De^{0}:  \,\mathrm{Im}\,[-\third (F^{(a)}+G^{(a)})+\third (F^{(b)}+G^{(b)})]
= 0 \,,\\ 
&& \De^{-}: -\,\mathrm{Im}\,[F^{(a)}+G^{(a)}]
= \mathrm{Im}\,\Si_\De'\,,\nn
\eea
where 
$\Si_\De'=\pa/\pa \slap \, \Si_\De(\slap)|_{\slap=m_\De}$, and hence
$\mathrm{Im}\,\Si_\De'= 4\pi C^2 (-\la^2 + \frac{7}{3}\la^3)$.

At the same time, the `naive' minimal-substitution procedure
[\Eqref{minsub}], that happens to
work for the muon, fails here miserably. It would predict that
the magnetic moment contribution would go with the same
power as the self-energy [\Eqref{coin}], which for the absorptive part means 
$\mathrm{Im}\,\mu 
\sim \mathrm{Im}\,\Si(m_\De)\sim \la^3$. In reality it goes as
$\la$. E.g., for the $\De^+$:  
\bea
\mathrm{Im}\, \mu_{\De^+} &= & (e/2m_\De)\, \mathrm{Im}
[\third F^{(a)}+
\mbox{$\frac{2}{3}$}F^{(b)}] \nn\\
&=& 
\mbox{$\frac{4}{3}$}\pi \, \mu_N C^2 \,\la
+ O(\la^2).
\eea
The fact that the self-energy goes as $\la^3$, while
$\mathrm{Im}\,\mu$ as $\la$
has as a consequence the enhancement of the lifetime
change in the magnetic field by two powers of $\la$.  

Quantitatively such enhancements of the lifetime change over the lifetime
by the phase-space volume do not make much difference in the above examples.
However, it shows that it might be useful to look for manifestations of the lifetime
change in the medium where the phase-space volume can be varied.

\section{Conclusions and outlook} 
We have examined her a concept of the `absorptive
magnetic moment'  --- an intrinsic property of an unstable particle,
together with the width or the lifetime. It manifests itself
in the change of the particle's lefetime in an external magnetic field,
see \Eqref{fin} below.
We have computed this quantity  for
 the examples of muon, neutron and $\De$-resonance to leading order
in couplings.
In all the three cosidered cases the effect on the lifetime is tiny for normal magnetic
fields: 
in a uniform field of 1 Tesla
the change in the lifetime is of order of $10^{-13}$ percent, at most.

In the case of the muon we have computed this effect to the
leading order in the electroweak coupling;
the change in the lifetime is
\beq
\eqlab{fin}
\De \tau  =
- 2 \,\mathrm{Im} \,\mu \, B_z\, \tau^2
 =  - 96\pi^3 e B_z/(G_F^2 m^7)\,,
\eeq
or, numerically, $|\De\tau| \lesssim 
6\times 10^{-21} \, (B/\mathrm{T}) \,\mathrm{sec}$.
A direct measurement of this effect is therefore 
 beyond the present experimental precision.  
Nevertheless, it is worthwhile to investigate the effect of the magnetic field
on the differential decay rates, with the hope that some asymmetries
could show a significantly bigger sensitivity.

A notable feature of this effect is that the
relative change of the lifetime 
is inversely proportional to the phase space.
It goes as $(m_\mathrm{n}-m_\mathrm{p})^{-1}$ 
in the neutron case, and as $(m_\De-m_N)^{-2}$ in the $\De$-resonance case.
(The difference in power is apparently because the neutron decays
solely into fermions while the $\De$ has a
boson in the decay product.)     
One can expect that in the conditions where the phase-space
is significantly reduced, e.g.\ for the neutron 
in  nuclear medium, the effect
of the lifetime change may become measurable. 

Especially interesting would be to evaluate the manifestations of this effect
in neutron star formations. Not only the phase-space of the neutron decay
is shrinking, the protons decay too, and all that occurs in 
magnetic fields as large as $10^10$ Tesla. Even larger fields
can be achieved in atomic or nuclear systems.  
Finally, it is worthwhile to point out that in lattice QCD studies strong
magnetic fields are standardly used to compute the electromagnetic 
properties of hadrons. Combined with the lattice techniques of extracting 
the width, the relation between the absorprive part 
and the lifetime change may allow to compute the former on the lattice
for unstable hadrons. 


\section*{Acknowledgments}
We thank Barry Holstein and 
Marc Vanderhaeghen for a number of insightful discussions.
The work of V.P.\ is partially supported  by the European Community-Research Infrastructure Activity under the FP6 "Structuring the European Research Area" programme (HadronPhysics, contract RII3-CT-2004-506078).

\end{document}